\documentclass[prl,twocolumn,preprintnumbers,amsmath,amssymb,floatfix]{revtex4}
\usepackage{amsfonts}
\usepackage{bbold}
\usepackage{mathrsfs}
\usepackage{color}
\usepackage{epstopdf}
\usepackage{graphicx}
\usepackage{bm}
\usepackage{silence}
\usepackage{amssymb,dsfont}

\newcommand{\rr}{  {\mathbf r} }
\newcommand{\ee}{ {\mathbf e} }

\newcommand{\BEQ}{\begin{equation}}
\newcommand{\EEQ}{\end{equation}}
\newcommand{\BEA}{\begin{eqnarray}}
\newcommand{\EEA}{\end{eqnarray}}

\newcommand{\mttp}[1]{\textcolor{black}{#1}}

\begin{document}

\title{
Relation between heterogeneous frozen regions in supercooled liquids \\
and non-Debye spectrum in the corresponding glasses
}


\author{Matteo Paoluzzi$^1$}
\email{Matteo.Paoluzzi@roma1.infn.it}                                  

\author{Luca Angelani$^{1,2}$}
\author{Giorgio Parisi$^{1,3,4}$}
\author{Giancarlo Ruocco$^{1,5}$}

\affiliation{
$^1$ Dipartimento di Fisica, Sapienza Universit\`a di Roma, Piazzale A. Moro 2, I-00185, Rome, Italy \\
$^2$ ISC-CNR,  Institute  for  Complex  Systems,  Piazzale  A.  Moro  2,  I-00185  Rome,  Italy \\
$^3$ Nanotec-CNR, UOS Rome, Sapienza Universit\`a di Roma, Piazzale A. Moro 2, I-00185, Rome, Italy \\
$^4$ INFN-Sezione di Roma 1, Piazzale A. Moro 2, I-00185, Rome \\
$^5$ Center for Life Nano Science, Istituto Italiano di Tecnologia, Viale Regina Elena 291, I-00161, Rome, Italy
}

\begin{abstract}
Recent \mttp{numerical studies on glassy systems provide evidences for a population of non-Goldstone modes (NGMs)
in the low-frequency spectrum of the vibrational density of states $D(\omega)$. Similarly to Goldstone modes (GMs), i. e., phonons in solids, 
NGMs are soft low-energy excitations. However, differently from GMs, NGMs are localized excitations.}
Here we first show that \mttp{the parental temperature $T^*$ modifies the GM/NGM ratio in $D(\omega)$. In particular, 
the phonon attenuation is reflected in a parental temperature dependency of the exponent $s(T^*)$ in the low-frequency
power law $D(\omega) \sim \omega^{s(T^*)}$, with $2 \leq s(T^*) \leq 4 $.}
%
Secondly, by comparing $s(T^*)$ with $s(p)$, \mttp{i. e.,} the same quantity 
obtained by pinning \mttp{a} $p$ particle \mttp{fraction}, we suggest that $s(T^*)$ reflects the
presence of dynamical heterogeneous regions of size $\xi^3 \mttp{\propto} p$. Finally, we 
provide an estimate of $\xi$ \mttp{as a} function of $T^*$, finding a mild power law 
divergence, $\xi \sim  (T^* - T_d)^{-\alpha/3}$, 
\mttp{with }$T_d$ \mttp{the dynamical crossover temperature and} 
$\alpha$ \mttp{falling} in the range $\mttp{\alpha \in }[0.8,1.0]$.
 \end{abstract}

\maketitle

\paragraph*{Introduction.}\label{Introduction}

At small enough {\textcolor{black}{frequencies}} $\omega$, the density of states $D(\omega)$ of \mttp{a} three 
dimensional \mttp{glass}
follows Debye's law $D(\omega) \sim \omega^2$. This is because {\textcolor{black}{at large enough length scale}} glasses are continuum media and thus phonons dominate the low frequency spectrum \cite{kittel2005introduction}.  However, compared with crystals, glasses show thermodynamic anomalies at low temperatures.  For instance, the thermal conductivity $\kappa(T)$ scales with $T^2$ \cite{Exp1} instead $T^3$, as predicted by Debye's law \cite{kittel2005introduction}. Moreover, also the specific heat $C_v$ below $1$ K deviates from Debye's law acquiring a linear dependency on $T$ \cite{Exp1}. Remarkably, these anomalies are shared by a broad class of glassy systems providing evidences of universality.

As it has been noticed in Ref. \cite{Gurarie03}, in disordered media and small $\omega$, $D(\omega)$  takes contributions from both, extended Goldstone bosons, e. g., phonons in structural glasses or spin-waves in Heisenberg spin glasses, and non-Goldstone modes, i. e., excitations that are not generated by the spontaneous symmetry breaking of a continuous symmetry. The Goldstone contribution gives rise to the Debye spectrum ${\textcolor{black}{D(\omega)}} \sim \omega^{d-1}$, with $d$ the number of spatial dimensions. The non-Goldstone sector is still soft, i.e., normal modes whose density of states vanishes as a power law ${\textcolor{black}{D(\omega)}}\sim \omega^{s}$ \cite{Gurarie03}, but it is populated by localized modes.
Only in the last few years, thanks to the possibility 
of eliminating Goldstone bosons from the low-energy spectrum \cite{Baity-Jesi15,Lerner2016} 
or discriminating non-extended modes from the extended ones \cite{Mizuno14112017},
it has been possible to observe numerically the non-Goldstone sector in numerical simulations 
%
obtaining that, in agreement with 
arguments suggested in Ref. \cite{Gurarie03}, 
non-Goldstone modes give  a contribution to $D(\omega)$ that scales with $s=4$.


In a previous work, 
we showed that a population of soft-localized modes with 
$\mttp{D(\omega)} \sim \omega^{s(p)}$ and $2\leq s(p) \leq 4$ emerges in the low-frequency spectrum of a three dimensional model of glass when a fraction
$p$ of particles are randomly frozen \cite{pnas}. In particular, the value of the \textcolor{black}{effective} exponent $s(p)$ {\textcolor{black}{starts from $s$=2 at $p$=0 and}} approaches $4$ above a threshold $p_{th}$ value that is of the order of $50 \%$ of frozen particles. 

\mttp{In this paper we study the properties of the vibrational density of states of a model of glass in its inherent states, i. e., configurations that
minimize the potential energy at $T=0$. Inherent states have been obtained after a fast quench from equilibrium configurations at the parental
temperature $T^*$.}
%
%
In agreement with Ref. \cite{Lerner_rapid}, the slope \mttp{in} the tail of $D(\omega)$ depends on the parental temperature $T^*$.
Moreover, as  a first result, we observe a progressive attenuation of the Debye spectrum in favor of the non-Debye one as $T^*$ approaches {\textcolor{black}{from above}} the dynamical crossover temperature $T_d$.

\mttp{It turns out that the Debye scaling $s$=2 holds at $T^* \gg T_d$, $s$ increases by decreasing $T^*$, and it saturates}  to $s=4$ right {\textcolor{black}{above}} $T_d$.
This crossover between Debye to non-Debye is accompanied by a progressively localization of the normal modes below the Boson peak. 
\mttp{We will show that it is possible to relate the suppression of extended excitations
with the proliferation of spatially heterogeneous regions.}
\mttp{We observe that} the behavior of $s(T^*)$ mirrors that of $s(p)$ observed in \mttp{Ref.} \cite{pnas}, indicating an increase in \mttp{the} size of frozen heterogeneous regions 
\mttp{as} $T^*$ \mttp{decreases.} 
\mttp{This finding suggests } 
a way \mttp{for measuring}
the size of these regions.

%
\mttp{In particular, we are able to perform }
a mapping between 
the properties of the inherent states of the randomly pinned system at high parental temperatures, i. e., $(T^*=\infty,p)$, with the inherent structures of the same system at low temperatures without frozen particles. i. e., $(T^*,p=0)$. 
\mttp{Defining $s(T) \equiv s(T^*,p=0)$ and $s(p) \equiv s(T^*=\infty,p) $ and}
looking at the solution of $s(T^*) \!=\! s(p)$,
we \mttp{will} show that the resulting curve $p(T^*)$ provides an estimate for
\mttp{a} correlation length $\xi$, being $\xi \equiv \xi_{pin} =(pN/\rho)^{1/3}$.
We are then able to extract  the behavior of $\xi_{pin}$ as a function of \mttp{$T^*$}.
 It turns out that $\xi_{pin}$ is compatible with a power law divergence at $T_d$, $\xi_{pin}^3 = (T^*-T_d)^{-\alpha}$ and $\alpha$ not far from one.

\mttp{It is worth noting that pinned particles 
have been intensively employed for gaining insight into the structural and dynamical
properties of glassy materials in both, analytical models \cite{cammarota2012ideal_pin,franz2006metastable,Szamel} and numerical simulations \cite{sch,biroli2008thermodynamic,ozawa2015equilibrium_pin,kob2012non_pin,brito2013jamming_pin,nagamanasa2015direct_pin,karmakar2013random,chakrabarty2015dynamics,PhysRevE.90.052305,PhysRevLett.110.245702,kim2003effects}. 
Here, in parallel with the study of $D(\omega)$ as a function of $T^*$, 
we use the method of random pinning to gain insight into the $T^*$ dependence of the growing correlation lengths in structural glasses through the low-frequency spectrum of $D(\omega)$.}

%

\paragraph*{Methods.}\label{Methods}
We consider a standard $50\!:\!50$ binary mixture composed of $N\!=\!N_A\!+\!N_B$ spherical particles at density $\rho\!=\!N/L^3\!=\!1$  \cite{Bernu87,Grigera01,PRX_Berthier}.
The system is  enclosed in a cubic box of side $L$ where periodic boundary conditions are considered.
Particle radii are $\sigma_A$ and $\sigma_B$ with $\sigma_A/\sigma_B\!=\!1.2$ 
and $\sigma_A\!+\!\sigma_B\!\equiv\! \sigma \!=\!1$ \cite{Grigera01}. 
\mttp{Details about numerical simulations can be found in Supplemental Material (SM)}. 
We consider hybrid Brownian/Swap Monte Carlo simulations that combine the numerical integration of the equations of motion with Swap Montecarlo moves \cite{Grigera01}.
For computing dynamical properties, i.e., the dynamical temperature $T_d$, 
 the 4-point susceptibility
$\chi_4(t)$ \cite{Lacevic2003,Biroli_2004}, and the 4-point correlation $S_4(q,t)$ \cite{Kob1997,Lacevic2003,PhysRevLett.112.097801,flenner2015fundamental}, 
we consider the Brownian evolution of thermodynamically stable configurations
obtained through hybrid Swap/Brownian dynamics \mttp{for system sizes $N=10^3,20^3$}.
%
%
\mttp{The minimization of the mechanical energy has been performed through the Limited-memory Broyden-Fletcher-Goldfarb-Shanno
algorithm \cite{bonnans2006numerical}. We then compute the eigenvalues of the dynamical matrix, i. e.,  the  Hessian matrix $\mathbf{M}$,
and thus we obtain the spectrum of the harmonic oscillations around the inherent structure. 
}
%
%
We also considered configurations equilibrated at high temperatures, i. e., $T^* \gg T_d$, where
a finite number of particles $pN$, with $p \in[0,1[$, 
are maintained 
frozen during the minimization (see Ref. \cite{pnas} for details).  
The 
eigenfrequencies
are $\omega_\kappa^2=\lambda_\kappa$, \mttp{with $\lambda_\kappa$ the $\kappa-$th eigenvalue of $\mathbf{M}$}.
%
\begin{figure}[!t]
\centering
\includegraphics[width=.235\textwidth]{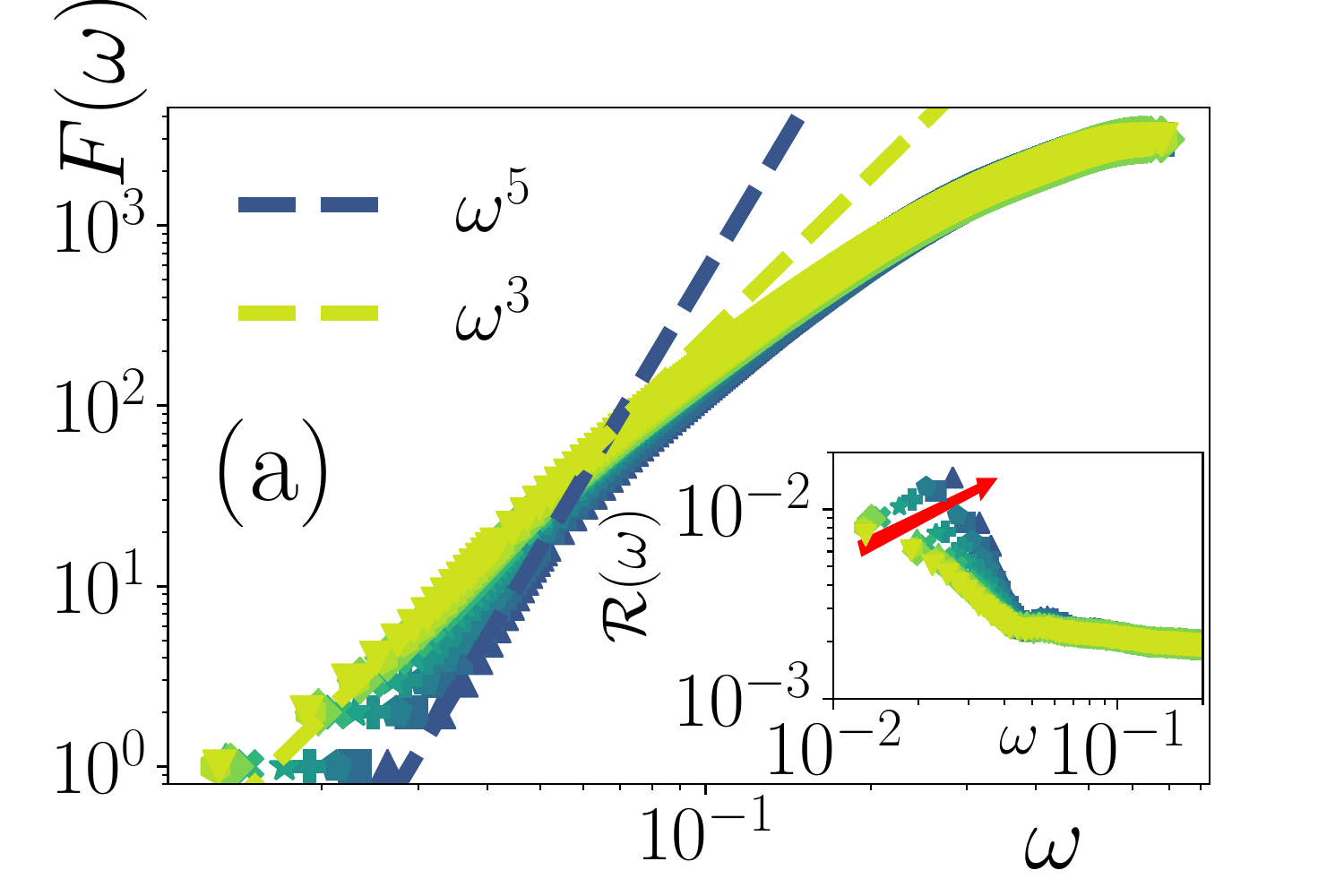}
\includegraphics[width=.235\textwidth]{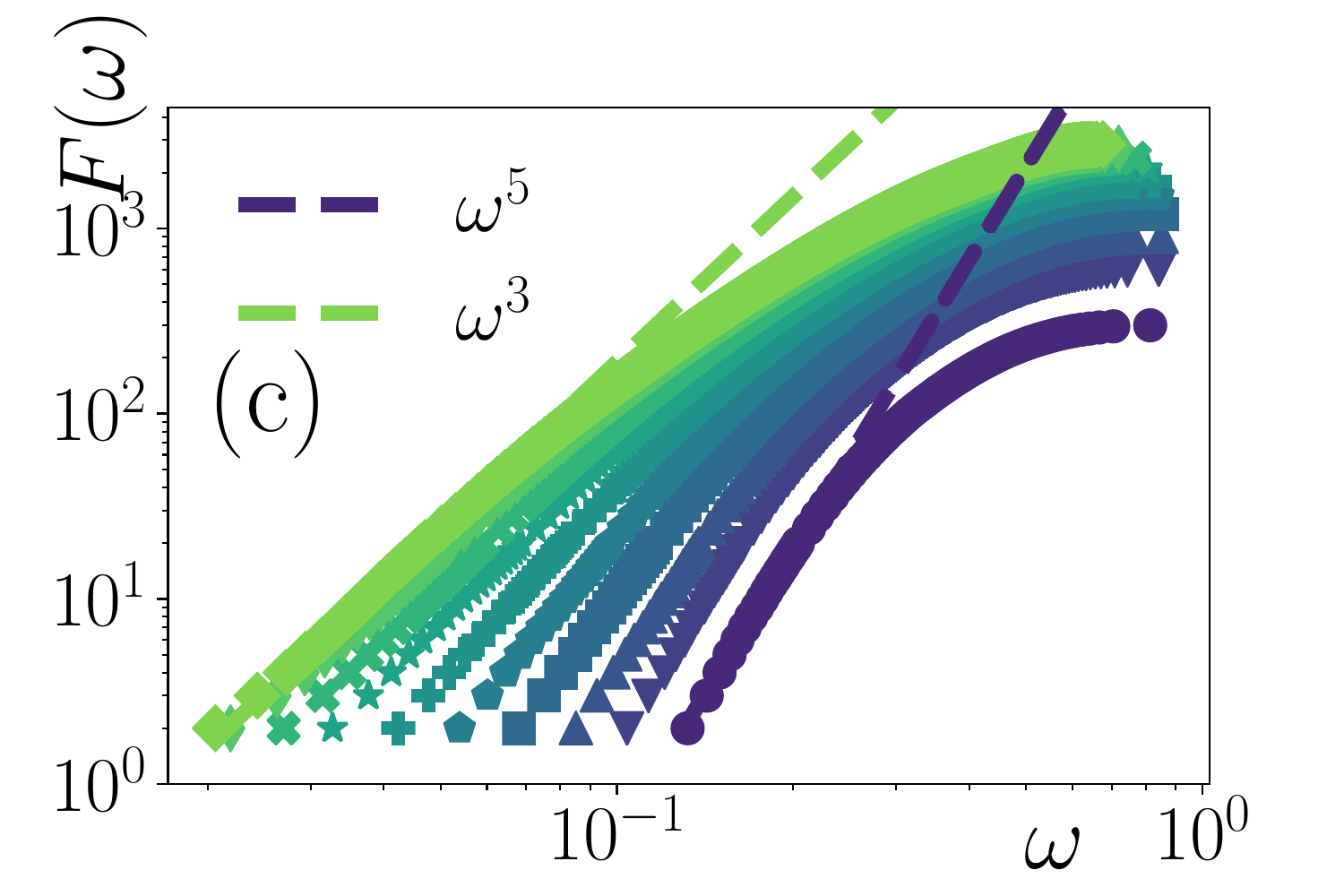} \\ 
\includegraphics[width=.235\textwidth]{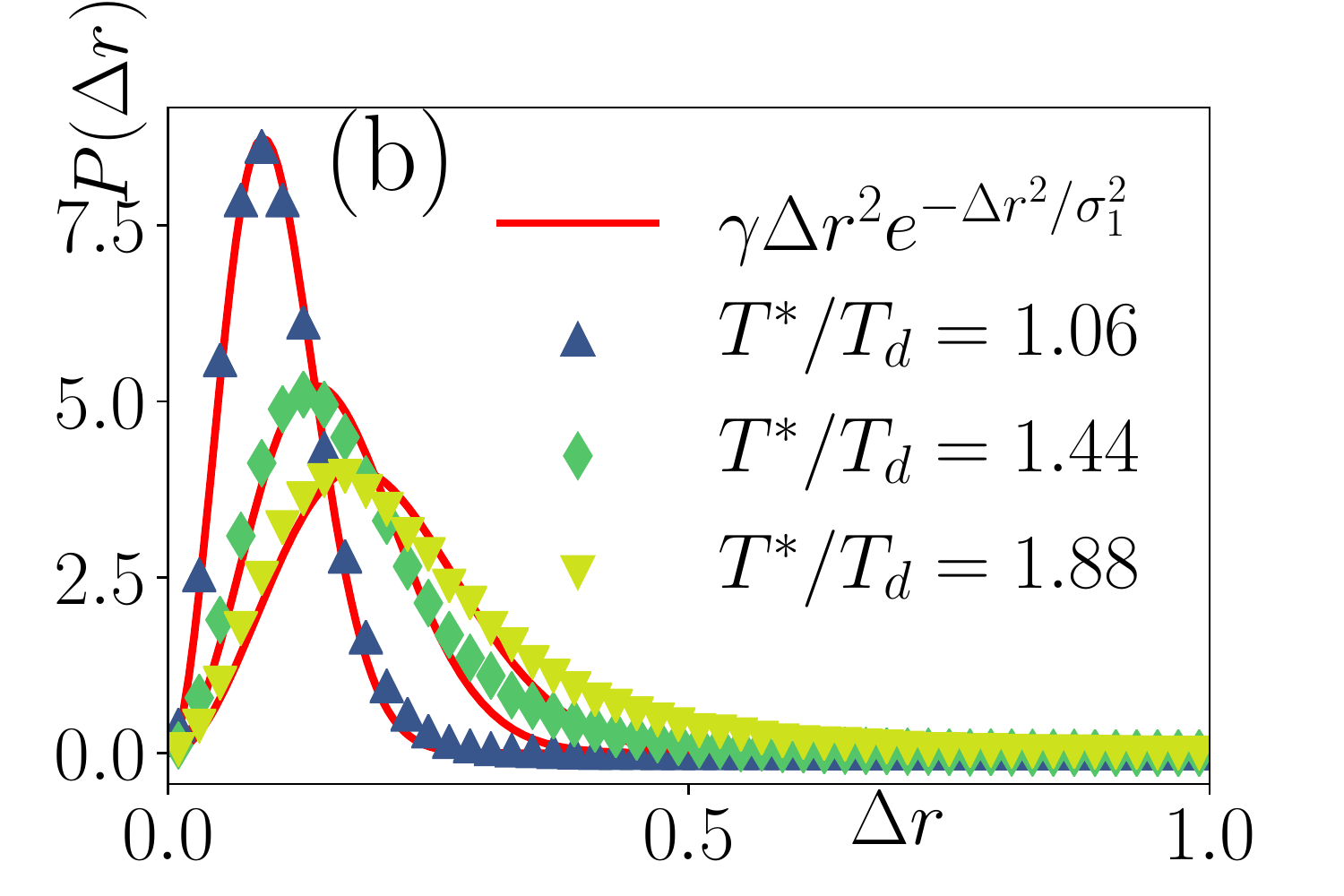}
\includegraphics[width=.235\textwidth]{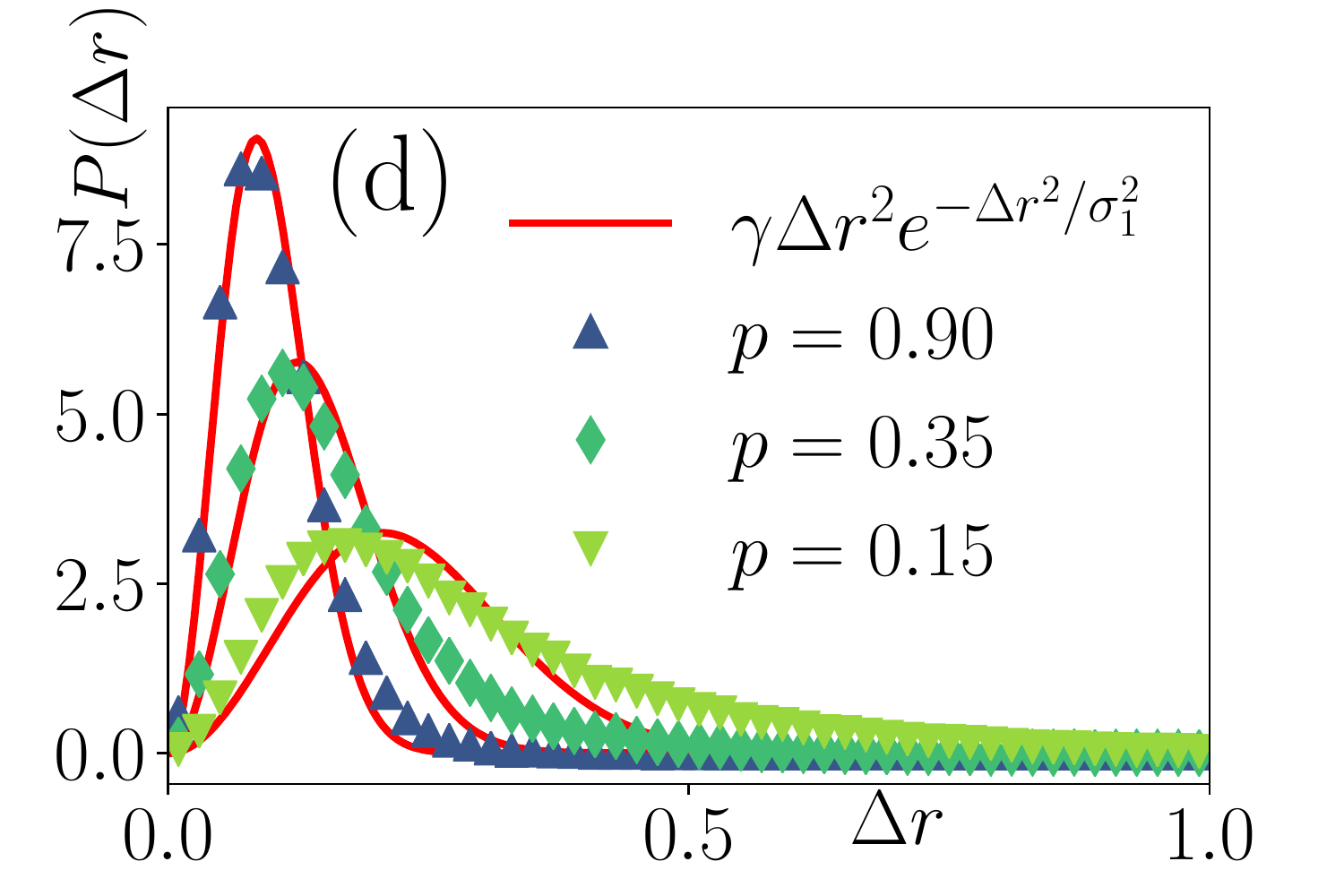}
\caption{ (a) Cumulative {\textcolor{black}{density of states}} $F(\omega)$ {\textcolor{black}{as function of the parental temperature} $T^*$ for $N=10^3$.}
\textcolor{black}{Temperatures decrease from yellow to blue, $T^*/T_d=1.87,1.56,1.50,1.44,1.38,1.31,1.25,1.19,1.13,1.06$.}
Inset: Inverse participation ratio ${\textcolor{black}{\mathcal{R}}}(\omega)$. 
(b) Probability distribution function $P(\Delta r)$ of the total displacement $\Delta r$ by varying temperature. 
(c)  Cumulative function $F(\omega)$ as the fraction of frozen particles $p$ increases at high temperature $T^*/T_d=3.12$.
\textcolor{black}{$p$ increases from green to violet, $p=0.05,0.1,0.2,0.3,0.4,0.5,0.6,0.7,0.8,0.9$.} 
(d) $P(\Delta r)$ at $T^*/T_d=3.12$ as $p$ increases from green to blue.
}\label{fig:fig1}
\end{figure}
We focus our attention on the cumulative $F(\omega)$=$\int_0^\omega d\omega^\prime \, D(\omega^\prime)$ of the density of states $D(\omega)$=$\mathcal{N}^{-1} \sum_\kappa \delta(\omega - \omega_\kappa)$, with $\mathcal{N}$  the number of non-zero modes. 
\mttp{We study the localization property of the normal-mode $\omega$ through the inverse of the participation ratio}
$\mathcal{R}(\omega) \equiv \sum_i | \ee_i (\omega )|^4  / \left( \sum_i | \ee_i (\omega )|^2 \right)^2$ 
where $\ee_i(\omega)$ is the eigenvector of the mode $\omega$ \cite{Bell70}. 
%
%
%
%
Let $\rr \! \equiv \! (\rr_1,...,\rr_N)$ be a configuration of the system thermalized at temperature $T^*$.
We indicate with $\rr^0 \! \equiv \! (\rr_1^0,...,\rr_N^0)$ the configuration that minimizes the mechanical 
energy.
\mttp{We also quantify the effect of the parental temperature on the inherent configuration through}
the distribution 
$P(\Delta r) = N^{-1} \sum_i \delta( \Delta r - \Delta r_i) $, with $\Delta r_i \equiv  |\rr_i- \rr_i^{0}|$, i. e., 
the total displacement covered by the particle $i$ for reaching the inherent configuration \mttp{$\rr^0$} starting from \mttp{$\rr$, i. e., the thermally} equilibrated one.
\mttp{Finally, we compare the behavior of $\xi_{pin}$ with the dynamical correlation length $\xi_{dyn}$ that is obtained by fitting 
the four-point correlation function $S_4(q,\tau_4)$ to an Ornstein-Zernike expression \cite{Kob1997}. 
The structural relaxation time $\tau_4$ has been evaluated looking at the peak of the $\chi_4(t)$ susceptibility \cite{lavcevic2003spatially,PhysRevLett.111.165701}. 
Details about the computation are provided in SM.}

\begin{figure}[!t]
\centering
\includegraphics[width=.45\textwidth]{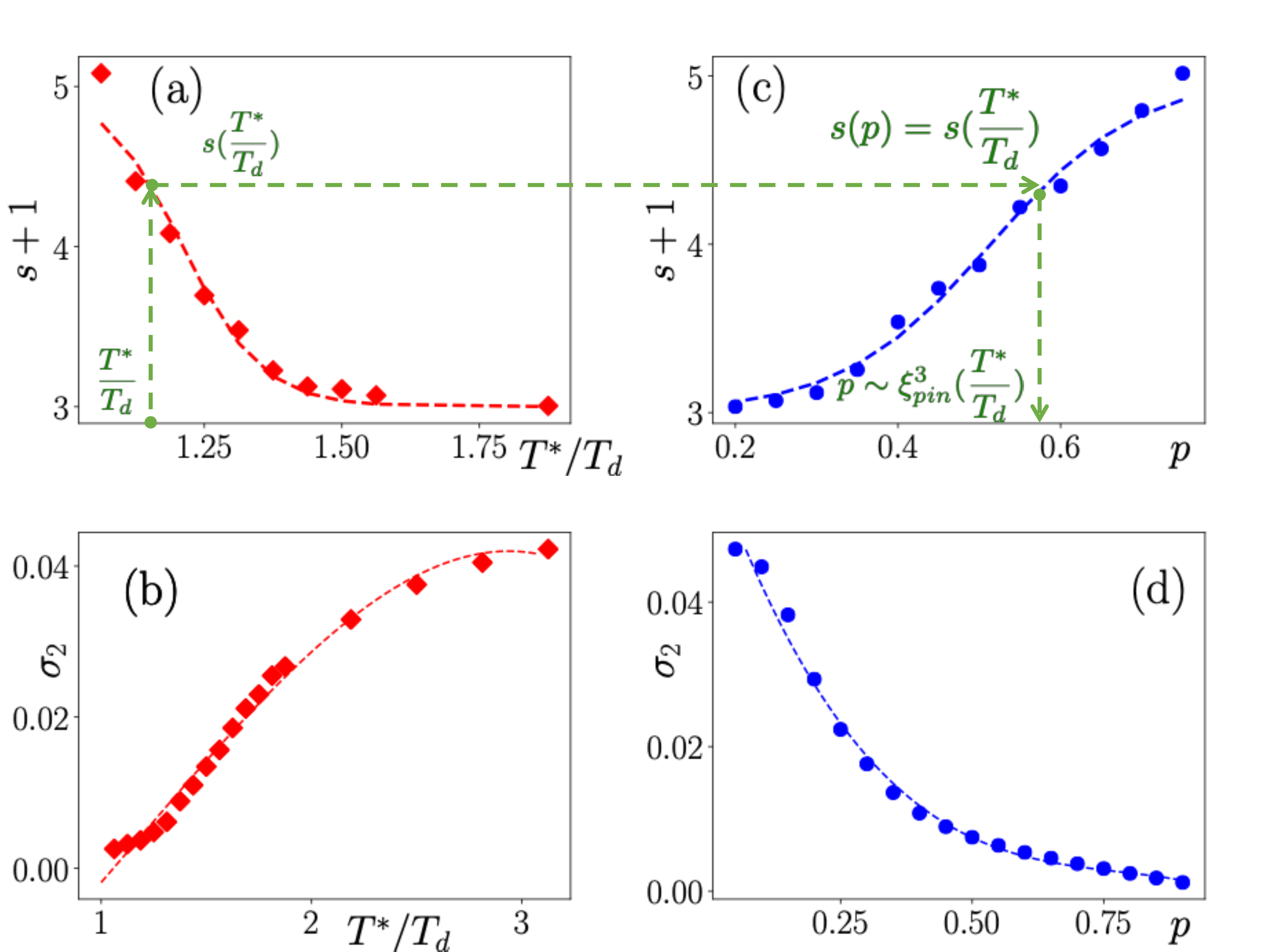}
\caption{ 
(a) Slope $s$ of the power law $F(\omega)\sim \omega^{s+1}$ as a function of the {\textcolor{black}{parental}} temperature $T^*$ for $N=10^3$.
(b) Variance of the distribution $P(\Delta r)$ as a function of $T^*$.
(c)  Slope $s$ as a function of $p$ at $T^*={\textcolor{black}{3.12}}$
(d) Variance of the distribution $P(\Delta r)$ as a function of $p$ at $T^*={\textcolor{black}{3.12}}$.
Dashed lines are guides to the eye. In panels (a) and (c) the green dashed arrows sketch the mapping employed for measuring $\xi_{pin}$.
}\label{fig:fig2}
\end{figure}

\paragraph*{Results.}\label{Results}
Let us start with discussing the effect of the parental temperature \mttp{$T^*$} on the cumulative
function. $F(\omega)$ is shown in \textcolor{black}{Fig. (\ref{fig:fig1}a)}  for different \mttp{$T^*$}
and system size $N=10^3$. 
Approaching the dynamical temperature, i.e.  $T^* / T_d \to 1$, the exponent of
the low-frequency power law  $F(\omega)\sim \omega^{s(T^*) + 1}$ increases
as temperature decreases departing from the Debye value $s=2$ to higher
values. A dependency of $s$ on both, the protocol adopted for cooling down the system 
and $T^*$, has been observed also in Ref. \cite{Lerner_rapid}. 
\textcolor{black}{As shown in Fig. (\ref{fig:fig2}a), $s+1\to5$ as $T^* \to T_d$.}
The exponents have been computed fitting the tail of $F(\omega)$ below the Boson peak \cite{buchenau1984neutron,binder2011glassy}
with a power law. Since the Boson peak is populated by extended modes, we select
the low-frequency sector through the ${\textcolor{black}{\mathcal{R}}}$ value of the mode $\omega$.
As one can see in the inset \textcolor{black}{of Fig. (\ref{fig:fig1}a)},
below $\omega \sim0.04$, $\mathcal{R}$ grows as frequency decreases. Moreover, in that region, $\mathcal{R}$ grows with decreasing \mttp{$T^*$} \textcolor{black}{(the arrow goes in the direction of decreasing temperatures)}, indicating
that low-frequency modes become more localized as temperature decreases.
The situation is different above $\omega \sim0.04$ where $\mathcal{R}$ approaches
the $1/N$ limit. 
The increasing in $\mathcal{R}(\omega)$ and the behavior $F(\omega)\sim \omega^{1+s(T^*)}$ on lowering frequency is consistent with the presence of soft-localized modes. 

In order to gain insight into the nature of the rearrangements made
by the system for reaching the {\textcolor{black}{inherent}} configuration $\rr^0$, we have then computed the distribution $P(\Delta r)$
that is shown in Fig. (\ref{fig:fig1}b). 
The distribution becomes peaked at {\textcolor{black}{lower and lower}} $\Delta r$ values as temperature decreases indicating that particles in configurations
at lower temperature turn to be more caged during minimization.  The red curves in Fig. (\ref{fig:fig1}b) are fits to $\gamma \Delta r^2 e^{-\Delta r^2 / \sigma_1^2}$, with $\gamma$ and $\sigma_1$  {\textcolor{black}{adjustable}} parameters. {\textcolor{black}{One can notice the presence of }} non-Gaussian tails  
at high temperatures {\textcolor{black}{that}} progressively disappear for $T^* \to T_d$. 
The non-Gaussian tails indicate that particles travel long distances  for reaching 
the optimal configuration when the parental configuration is taken at high $T^*$. As $T^*$
decreases towards $T_d$,  particles turn to be more caged by their neighbors and they thus perform small uncorrelated displacements
to find the closer inherent structure.
To be more quantitative, 
{\textcolor{black}{we have also computed the true variance \textcolor{black}{$\sigma_2$} of the distribution $P(\Delta r)$, $\sigma_2(T^*)$}}\textcolor{black}{(see Fig. (\ref{fig:fig2}b,\ref{fig:fig2}d))}.

{\textcolor{black}{A similar phenomenology is observed looking at the}} \textcolor{black}{system}  {\textcolor{black}{at high $T^*$ but including a}} fraction of frozen particle during the research of the inherent structure \cite{pnas}.
In particular,  when the concentration of frozen particle\textcolor{black}{s} is large enough, moving particles are caged by the
non-moving ones. 
In Fig. (\ref{fig:fig1}c) the cumulative $F(\omega)$ is shown at $T^*/T_d=3.12$ by varying the fraction of 
frozen particles $p$, that increases from 
\textcolor{black}{left to right}.
We have thus computed the distributions $P(\Delta r)$ when
a fraction $p$ of particles is maintained frozen and the results are shown in Fig. (\ref{fig:fig1}d). Again, the
red curves are the {\textcolor{black}{gaussian}} fit.  
The behavior of $P(\Delta r)$ with increasing
the fraction of frozen particles $p$ is qualitatively the same obtained with decreasing temperature \textcolor{black}{in the un-pinned system}.  
This is a strong indication that the same crossover from a spectrum dominated by soft-extended  modes 
to soft-localized modes takes place in both protocols, i.e, decreasing parental temperature or increasing the fraction of pinned particles. The advantage of introducing randomly frozen particles lies in the fact that controlling $p$ we are also controlling the typical size $\xi_{pin}$ of frozen regions. In order to define $\xi$, we notice that the number of frozen particles $N_p= p N$ is naturally proportional to the volume of frozen particles $V_p$,  thus $\xi^3 \propto p N$.

{\textcolor{black}{In order to make quantitative progresses, we look at the curves $s(T^*)$ and $s(p)$ obtained from $F(\omega)$, as well as at $\sigma_1(T^*)$ and $\sigma_1(p)$ (or, equivalently, at $\sigma_2(T^*)$ and $\sigma_2(p)$) obtained from $P(\Delta r)$. As already noticed $s(T^*)$ increases with decreasing $T^*$: this behavior is reported in Fig. (\ref{fig:fig2}a). The dashed-red line is a fit to logistic curve. Similarly, the behaviour of $\sigma_2(T^*)$ as a function of $T^*$ is reported in Fig. (\ref{fig:fig2}b).}} In panels (c) and (d) of the same figure we show the same observables 
as a function of the fraction of frozen particles $p$ for configurations thermalized {\textcolor{black}{well}} above the dynamical temperature, i. e., $T^*/T_d=3.12 $.

\begin{figure}[!t]
\centering
\includegraphics[width=.425\textwidth]{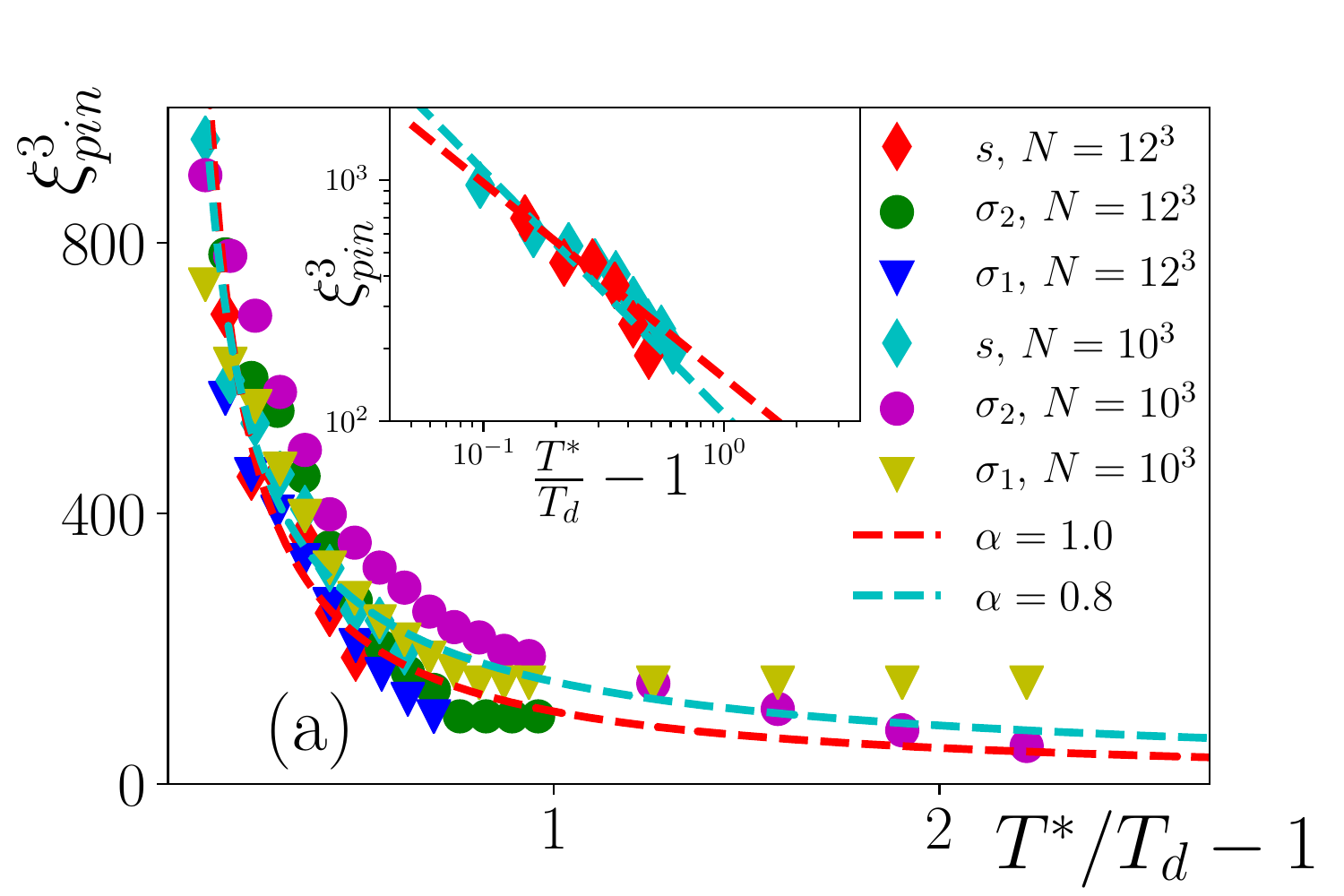} \\
\includegraphics[width=.425\textwidth]{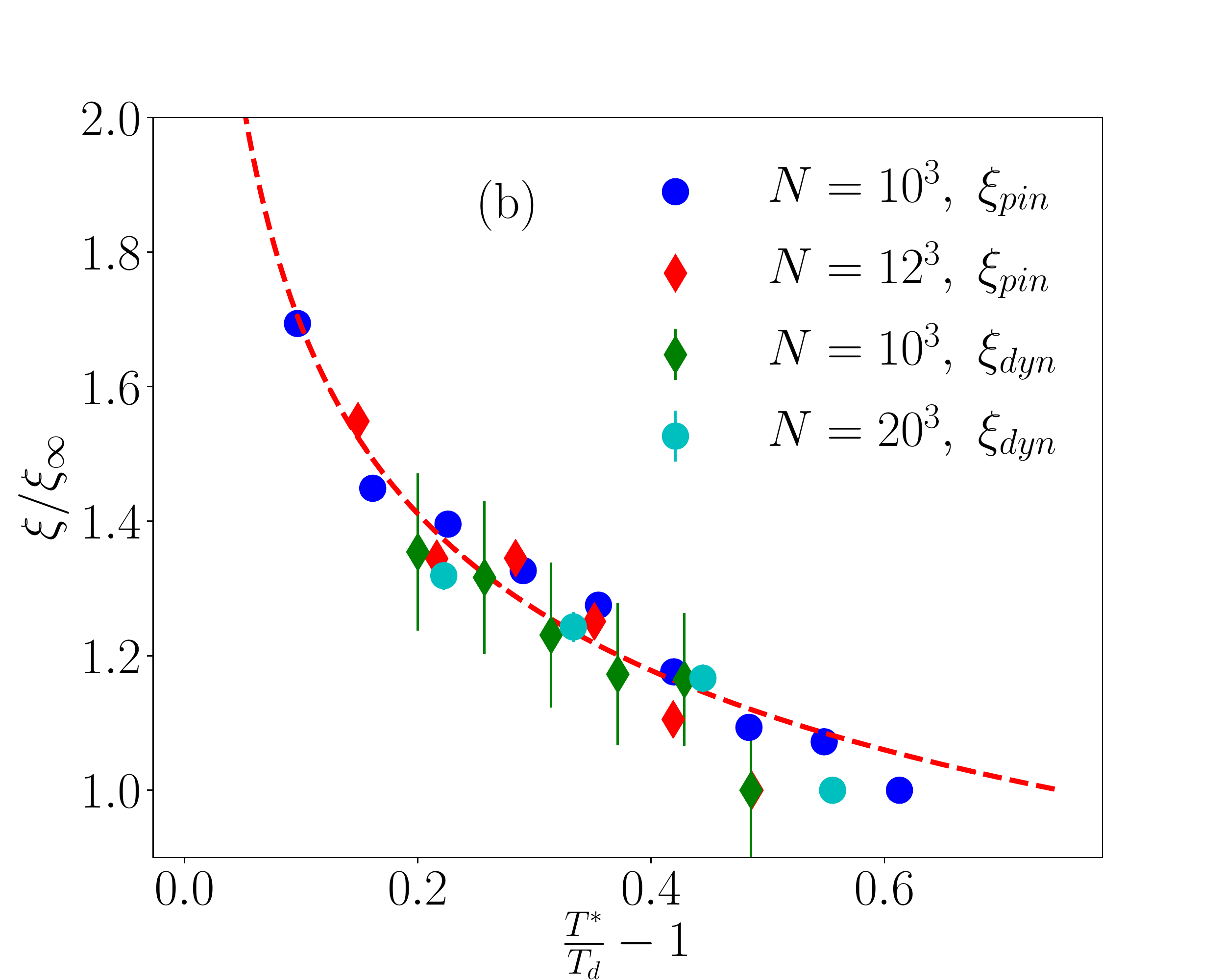}
\caption{ 
(a) Correlation length $\xi_{pin}^3$ defined in the main text  as a function of temperature $T^*$ and
estimated through different observables \textcolor{black}{for $N=10^3,12^3$}. Diamonds refer to $s(T^*)=s(p)$ method,
 circles to $\sigma_2(T^*)=\sigma_2 (p)$, and triangles using  $\sigma_1$, i. e., fitting $P(\Delta r)$ to
  $\gamma \Delta r^2 e^{-\Delta r^2 / \sigma_1^2}$ and thus considering $\sigma_1(T^*)=\sigma_1(p)$.
The inset highlights the behavior of $\xi_{pin}^3$
computed through the exponent $s$ for $N=10^3,12^3$, cyan and red symbols, respectively. 
Dashed lines are fits to the power law $(T^*-T_d)^{-\alpha}$ with $\alpha=0.8$ (red) and $\alpha=1.0$ (cyan).
(b) Comparison between the correlation length $\xi_{pin}$ 
 and the dynamic length $\xi_{dyn}$. $\xi_{\infty}$ indicates the value of $\xi$ at high temperatures.
 The red dashed line is the best fit $(T^*-T_d)^{-\alpha /3}$ with $\alpha\sim0.8$.
}
\label{fig:fig3}
\end{figure}

We can thus provide a quantitative estimate of the behavior of $\xi_{pin}$ as a function of $T^*$ mapping the properties of the pinned system
into the properties of the thermal system. In particular, we assume that, in a system
where a fraction $p$ of particles are frozen randomly in space, one introduces a correlation length
$\xi_{pin} \equiv \left( p N / \rho \right)^{1/3}$. For \textcolor{black}{inferring} the
correlation length $\xi_{pin}$ in the real system, i. e., without artificially frozen particles, 
we invert the relation $\mathcal{O}(T^*,p=0) = \mathcal{O}(T^*\gg T_d,p)$, 
where $\mathcal{O}$ is a generic observable, \mttp{and thus we obtain a function $p(T^*)$ that allows to measure $ \xi_{pin}^3(T^*)$}.
The green dashed arrows in Fig. (\ref{fig:fig2}a, \ref{fig:fig2}c) give a pictorial representation of the mapping we employ to infer $\xi_{pin}$ choosing as
an observable the exponent $s$.
The results of our analysis are shown in Fig. (\ref{fig:fig3})-a for system sizes $N=10^3,12^3$. Diamonds are
obtained considering the exponents of the power laws $s(T^*)$ and $s(p)$
as observable {\textcolor{black}{$\mathcal{O}$}} \mttp{for mapping}, i.e. $\mathcal{O} \equiv s$. Circles refer to the {\textcolor{black}{true}} variance of the distribution $P(\Delta r)$, {\textcolor{black}{$\mathcal{O} \equiv \sigma_2$}}. Ttriangles are obtained considering the {\textcolor{black}{parameter}} $\sigma_1$ from the fit of $P(\Delta r)$ to a gaussian distribution {\textcolor{black}{($\mathcal{O} \equiv \sigma_1$)}}.
The dashed curves are the power law $\xi_{pin}^3 \sim (T^*-T_d)^{-\alpha}$. 
The exponent $\alpha$ has been computed considering $\mathcal{O}=s$ and data set $N=10^3$ (cyan symbols), $N=12^3$ (red symbols). 
We then
obtain $\alpha_{max}=1.0$ and $\alpha_{min}=0.8$ for $N=12^3,10^3$, respectively, indicating that the exponent $\alpha$ varies in the range $\alpha\in[0.8,1.0]$. 
%
%
As one can appreciate, different observables $\mathcal{O}$ provide estimates for
$\xi$ that are consistent with the same mild power-law divergence as $T^*$ decreases towards 
$T_d$.

Dynamical heterogeneities are fingerprint patterns of glassy dynamics \cite{Berthier2011}. They 
suggest the existence of a dynamical correlation length $\xi_{dyn}$ that can be estimated through
multi-point correlation functions 
\cite{Biroli_2004,Biroli2006,lavcevic2003spatially,karmakar2009growing,PhysRevLett.111.165701,PhysRevLett.112.097801,flenner2015fundamental}.
\mttp{We thus compute $\xi_{dyn}$ for system sizes $N=10^3, 20^3$ and compare it with $\xi_{pin}$ computed before.}
The result is shown in Fig. (\ref{fig:fig3}b), the green circles are $\xi_{dyn}$ for $N=10^3$\mttp{, cyan circles refer to $N=20^3$}. \mttp{$\xi_{dyn}$ has been normalized
with the value of $\xi$ at high temperature}, i. e., $\xi_{\infty} \equiv \xi(T^*  \gg T_d)$.  As one can appreciate, 
both the correlation length $\xi_{pin}$ and  $\xi_{dyn}$ show a mild growth that is compatible with $(T^*-T_d)^{-1 /3}$.




\paragraph*{Discussion.}
 In this paper, we have explored the properties of the low-frequency excitations in a three-dimensional model glass {\textcolor{black}{obtained by fast quench from a well equilibrated super-cooled liquid configuration at $T=T^*$. The values of $T^*$ spanned from high temperatures down to}} the dynamical transition temperature $T_d$. 
We have shown that quasi-localized soft modes progressively populate the low-frequency spectrum. The density
of states of these {\it glassy modes} follows a scaling law $\omega^{s(T^*)}$ with
$2\leq s(T^*) \leq 4$. In particular, far away from the dynamical transition,
the low-frequency spectrum below the Boson peak is well described by Debye's law, i. e., 
$s(T^*)=2$. As $T^*$ decreases, $D(\omega)$ at small $\omega$ is still power law
with an exponent that is temperature dependent and deviates from Debye's law. In particular, 
$s$ starts \textcolor{black}{to increase} its value and $s(T^*) \to 4$ for $T^*\to T_d$. 
As shown here and also before in Ref. \cite{pnas}, the same {\textcolor{black}{quasi-}}logistic growth of $s$
is observed when, instead \textcolor{black}{of} varying the parental temperature, we introduce a fraction 
$p$ of frozen particles. In particular, the spectrum of the low-energy excitations below the Boson peak 
remains gapless and progressively deviates from Debye's law following a 
scaling $\omega^{s(p)}$, \textcolor{black}{with} $2\leq s(p) \leq 4$. In this case, $s(p)$ increases
as $p$ increases and $s\to 4$ above a threshold value $p_{th} \sim 0.5$. 

The emerging
phenomenology is consistent with a picture of heterogeneous regions where particles
experienced different mobilities \textcolor{black}{\cite{Kob1997,Glotzer,Berthier2011}.} \mttp{Considering a three dimensional system,} an estimate of the typical linear size $\xi$ of these
heterogeneous regions in the pinned system is provided  by $p^{1/3}$.
\mttp{This argument for the scaling of $\xi$ together with our estimate of $p(T)$ leads to an estimate of $\xi(T)$ that}  is compatible with the inhomogeneous Mode-Coupling theory
discussed in Ref. \cite{Biroli2006} where  $\xi_{dyn} \sim (T - T_d)^{-\nu}$ with $\nu=1/4$.

%

It has been shown in Refs. \cite{karmakar2012direct,gutierrez2015static} that
the statistical properties of the lowest eigenfrequency of  $D(\omega)$ 
can be employed to define 
 a static length scale whose behavior is consistent with 
the point-to-set  length \cite{biroli2008thermodynamic, PhysRevLett.111.165701,karmakar2009growing,PhysRevE.85.011102}.
Our study shows that \mttp{low-frequency modes in $D(\omega)$}
not only regulate the growing of a static length, and thus the changing in the thermodynamic properties of the system, but
also the growing of dynamic heterogeneous patterns.

In conclusion, \mttp{we showed that $D(\omega)$ provides useful information about both, the
structural properties of the glassy state through its inherent structures, and the dynamical properties of the corresponding
supercooled equilibrium configurations at $T^*$.}
%
\mttp{As a consequence,} from $D(\omega)$ we can extract important information about the correlation length 
of the heterogeneous regions in supercooled liquids, which, most likely, 
are the ultimate origin of the instability giving rise to the non-Goldstone modes \cite{schirmacher2006thermal,schirmacher2007acoustic,schirmacher2008vibrational,marruzzo2013vibrational,Marruzzo13,tomaras2013high,schirmacher2015theory}. 
A recent experiment showed that $D(\omega)$ results to be modified by natural hyperaging \cite{doi:10.1021/acs.jpclett.9b00003}.
As a future direction, it would be interesting \mttp{to investigate numerically $D(\omega)$ in the aging regime for understanding the relation between 
non-Debye spectrum and aging in glasses.}


\paragraph*{Acknowledgments.} 
GP acknowledges the financial support of the Simons Foundation (Grant No. 454949)
and the European Research Council (ERC)  (grant agreement No [694925]).
GP and MP were also supported by ADINMAT, WIS-Sapienza. 

\bibliography{glassybib}
\bibliographystyle{apsrev}
\end{document}